\documentclass[12pt]{article}
\title{Properties of Fermion Spherical Harmonics}

\author{\thanks{Author to whom correspondence should be addressed;
email: ghunter@yorku.ca}
Geoffrey Hunter and Mohsen Emami-Razavi\\
{\small\sl Centre for Research in Earth and Space Science},\\
{\small\sl York University, Toronto, Canada  M3J 1P3}}

\begin{document}

\maketitle

\vspace*{-0.25in}

\begin{abstract}
The Fermion Spherical harmonics [$Y_\ell^{m}(\theta,\phi)$ for half-odd-integer $\ell$ and $m$ - presented in a previous paper]
are shown to have the same eigenfunction properties as the well-known Boson Spherical Harmonics [$Y_\ell^{m}(\theta,\phi)$ for integer $\ell$ and $m$].
The Fermion functions are shown to differ from the Boson functions in so far as the ladder operators $M_+$ ($M_-$) that ascend (descend) the
sequence of harmonics over the values of $m$ for a given value of $\ell$, do not produce the expected result {\em in just one case}:
when the value of $m$ changes from $\pm\frac{1}{2}$ to $\mp\frac{1}{2}$; i.e.~when $m$ changes sign; in all other cases the ladder
operators produce the usually expected result including anihilation when a ladder operator attempts to take $m$ outside the range: $-\ell\le m\le +\ell$.

The unexpected result in the one case does not invalidate this scalar coordinate representation of spin angular momentum, because the
eigenfunction property is essential for a valid quantum mechanical state, whereas ladder operators relating states with different eigenvalues are
not essential, and are in fact known only for a few physical systems; that this coordinate representation of spin angular momentum differs from the
abstract theory of angular momentum in this respect, is simply an interesting curiosity.
This new representation of spin angular momentum is expected to find application in the theoretical description of physical systems and experiments
 in which the spin-angular momentum (and associated magnetic moment) of a particle is oriented
in space, since the orientation is specifiable by the spherical polar angles, $\theta$ and $\phi$.
\end{abstract}
PACS 03.65-w

\section{Introduction}
Our previous derivation and presentation of the Fermion Spherical Harmonics (i.e.~$Y_\ell^{m}(\theta,\phi)$ for $\ell$ and $m$ half odd-integers)
 \cite{Hunteretal} was motivated by the idea that the half-odd-integer
spherical harmonics could be a useful representation of particle spin; in particular for the electron
its spin is associated with a
magnetic moment having an orientation in space specifiable by the spherical polar angles $\theta$ and $\phi$, and
the two
Fermion spherical harmonics for $\ell=\frac{1}{2}$ and $m=\pm \frac{1}{2}$ correspond to the experimentally established spin states of the electron.
This explicit dependence upon the orientation in space of the electron's magnetic moment is a distinct advantage over
the commonly used abstract spin eigenfunctions  $\alpha$ and $\beta$ \cite[\S10.1,p.282]{Levine},
because the latter have no dependence upon any coordinates.

The investigation reported here was initiated by the observation that the Fermion Spherical Harmonics \cite{Hunteretal}
are not always inter-related by the well-known ladder operators \cite[\S5.4,p.115]{Levine}.

Our objectives in this research were threefold:
\begin{itemize}
\item[a)] to confirm that these functions of $\theta$ and $\phi$ are
eigenfunctions of the angular momentum operators, ${\bf M_z}$ and
${\bf M^2}$, with the expected eigenvalues,\footnote{The notation ${\bf M_z}$, ${\bf M^2}$, is taken
from Levine \cite[p.116,\P\,1]{Levine} as denoting {\em any} kind of angular momentum,
whereas spin angular momentum is usually denoted by ${\bf S_z}$, ${\bf S^2}$,
orbital angular momentum by
${\bf L_z}$, ${\bf L^2}$, and total angular momentum by ${\bf J_z}$, ${\bf J^2}$.}
\item[b)] to determine to what extent the ladder operators operating upon these
functions produce the usual, expected results (since they do fail in some cases),
\item[c)] to consider if there are reasons (failure of the ladder operators or otherwise)
that make these functions unsatisfactory
as a representation of spin angular momentum.
\end{itemize}

\subsection{Angular Momentum Operators}
The following definitions of the cartesian-components of the angular
momentum operators (in spherical polar angles,
$\theta,\phi$) are taken from  McQuarrie \cite[eq.6-85, page 217]{McQuarrie}
\begin{eqnarray}
{\bf M_x} &=& \frac{\hbar}{i}\left(-\sin\,\phi\,{\frac{\partial}{\partial\theta}}\, -
\cot\,\theta\,\cos\,\phi\,{\frac{\partial}{\partial \phi}}\right)\\
{\bf M_y} &=& \frac{\hbar}{i}\left(\cos\,\phi\,{\frac{\partial}{\partial\theta}} -
\cot\,\theta\,\sin\,\phi\,
{\frac{\partial}{\partial\phi}}\right)\\\label{defLz}
{\bf M_z} &=& \frac{\hbar}{i}\frac{\partial}{\partial \phi}
\end{eqnarray}
From these definitions the form of the operator for the square of the total
angular momentum, ${\bf M^2}$  is  \cite[eq.6-45, page 207]{McQuarrie}.
\begin{eqnarray}\label{defL2}
{\bf M^2} = {\bf M_x^2} + {\bf M_y^2} + {\bf M_z^2} = -{\hbar}^{2}\, \left(\!
{\frac{\partial^{2}}{\partial\theta^{2}}}\, + \cot\,\theta\,{\frac{\partial}{\partial\theta}}\, +
{\frac{1}{\sin^2\theta}{\frac{\partial^{2}}{\partial\phi^2}}} \right)
\end{eqnarray}

\subsection{Ladder Operators}
The Ladder Operators, ${\bf M_+}$  and ${\bf M_-}$ are defined by (\cite[p.116]{Levine}):
\begin{eqnarray}\label{MPM}
{\bf M_+} = {\bf M_x} + i\,{\bf M_y} \quad\quad\quad
{\bf M_- } = {\bf M_x} -  i\,{\bf M_y}
\end{eqnarray}
whose form in terms of $\theta$ and $\phi$ is \cite[p.182]{MerzbacherBook}:
\begin{eqnarray}\label{ladderopP}
{\bf M_+} = \hbar\,e^{(i\,\phi)}
\left(\frac{\partial}{\partial\theta} +i\,\cot\theta \,\frac{\partial}{\partial\phi}\right)\quad\quad
{\bf M_-} = \hbar\,e^{(-i\,\phi)}
\left(\frac{\partial}{\partial\theta} - i\,\cot\theta \,\frac{\partial}{\partial\phi}\right)
\end{eqnarray}
which are the appropriate forms for our purposes.

\section{Spherical Harmonics}
It is true in general (for integer and half-odd integer values of $\ell$ and $m$)
that a spherical harmonic has the form:
\vspace{-\baselineskip}
\begin{eqnarray}\label{genSH}
Y_\ell^{m} = e^{(i\,m\,\phi)}\,(\sin\theta)^{|m|}\,P_\ell^{|m|}(\cos\theta)
\end{eqnarray}
where
$P_\ell^{|m|}(\cos\theta)$ is a polynomial in $\cos\theta$ of order $(\ell-|m|)$
(this order is {\em always} an integer even when $\ell$ and $m$ are half-odd-integers).
Some of the Legendre functions are shown explicitly in Table 1 of our previous paper \cite[p.796]{Hunteretal};
from this table the first twelve Fermion spherical harmonics have the explicit (unnormalized) forms:

\begin{eqnarray}\label{Y1212}
Y_{\frac{1}{2}}^{\frac{1}{2}}  = \sqrt{\sin\theta}\,e^{ (\frac{1}{2}\,i\,\phi)}                             &\quad&
Y_{\frac{1}{2}}^{-\frac{1}{2}} = \sqrt{\sin\theta}\,e^{-(\frac{1}{2}\,i\,\phi)}                            \\\label{Y3232}
Y_{\frac{3}{2}}^{\frac{3}{2}}  = (\sin\theta)^{\frac{3}{2}}\,e^{ (\frac{3}{2}\,i\,\phi)}                 &\quad&
Y_{\frac{3}{2}}^{-\frac{3}{2}} = (\sin\theta)^{\frac{3}{2}}\,e^{-(\frac{3}{2}\,i\,\phi)}                \\\label{Y1232}
Y_{\frac{3}{2}}^{\frac{1}{2}}  = \cos\theta\sqrt{\sin\theta}\,e^{ (\frac{1}{2}\,i\,\phi)}               &\quad&
Y_{\frac{3}{2}}^{-\frac{1}{2}} = \cos\theta\sqrt{\sin\theta}\,e^{-(\frac{1}{2}\,i\,\phi)}              \\\label{Y5252}
Y_{\frac{5}{2}}^{\frac{5}{2}}  = (\sin\theta)^{\frac{5}{2}}\,e^{ (\frac{5}{2}\,i\,\phi)}                  &\quad&
Y_{\frac{5}{2}}^{-\frac{5}{2}} = (\sin\theta)^{\frac{5}{2}}\,e^{-(\frac{5}{2}\,i\,\phi)}                 \\\label{Y3252}
Y_{\frac{5}{2}}^{\frac{3}{2}}  = \cos\theta\,(\sin\theta)^{\frac{3}{2}}\,e^{ (\frac{3}{2}\,i\,\phi)}  &\quad&
Y_{\frac{5}{2}}^{-\frac{3}{2}} = \cos\theta\,(\sin\theta)^{\frac{3}{2}}\,e^{-(\frac{3}{2}\,i\,\phi)}  \\\label{Y1252}
Y_{\frac{5}{2}}^{\frac{1}{2}}  = (1\!-\!4\,\cos^2\!\theta)\sqrt{\sin\theta}\,e^{ (\frac{1}{2}\,i\,\phi)} &\quad&
Y_{\frac{5}{2}}^{-\frac{1}{2}} = (1\!-\!4\,\cos^2\!\theta)\sqrt{\sin\theta}\,e^{-(\frac{1}{2}\,i\,\phi)}
\end{eqnarray}

\section{The Eigenfunction Properties}

        In this section we {\em explicitly} confirm that the Fermion Spherical Harmonics are
eigenfunctions of  ${\bf M_z}$ and ${\bf M^2}$ with the expected eigenvalues.  This was {\em implicit} in
the derivation of these functions presented in our previous paper \cite{Hunteretal}.

\subsection{${\bf M_z}$ Operator}
From the general form of the Spherical Harmonics (\ref{genSH}) it follows immediately that
$Y_\ell^m$ is always an eigenfunction of ${\bf M_z}$ (\ref{defLz}) with eigenvalue $m\hbar$:
\begin{eqnarray}\label{proofLz}
M_z Y_\ell^m = \frac{\hbar}{i}\,\frac{\partial Y_\ell^m}{\partial\phi} =
\frac{\hbar}{i}\,(\sin\theta)^{|m|}\,P_\ell^{|m|}(\cos\theta)\,\frac{\partial e^{(i\,m\,\phi)}}{\partial\phi}
= m\,\hbar\,Y_\ell^m
\end{eqnarray}
and hence this eigenfunction property is proven for all cases
and requires no further consideration; the eigenvalue, $m\hbar$, is the physically expected value for the $z$-component of angular momentum both for integral and half-odd-integral values of $m$.

\subsection{${\bf M^2}$ Operator}
One conclusion from the general form of the Spherical Harmonics (\ref{genSH}) and from the
form of the ${\bf M^2}$ operator (\ref{defL2}) is that operation on
$Y_\ell^{+|m|}$ produces the same result as on
$Y_\ell^{-|m|}$; i.e.~the same result for the two harmonics that differ only in the sign of $m$.
This occurs because the only differentiation with respect to  $\phi$ in ${\bf M^2}$ is the
second derivative with respect to $\phi$, and in view of the simple exponential dependence of every harmonic on $\phi$, (\ref{genSH}), differentiation of this exponential factor twice produces the same result for positive $m$ as for negative $m$; in the latter case the multiplying negative sign introduced by the first differentiation is canceled by the second differentiation.  Thus it is only necessary to consider operation of the ${\bf M^2}$ operator on $Y_\ell^{+|m|}$.

\subsubsection{Some Special Cases}\label{12SC}
In view of the polynomial dependence of $Y_\ell^m$ on $\cos\theta$, it isn't possible to infer
the general result as we did for ${\bf M_z}$; nevertheless some especially simple cases and
the following six examples indicate that $Y_\ell^m$
is an eigenfunction of  ${\bf M^2}$ with the physically expected eigenvalue for both integral and
half-odd-integral values of $\ell$.  These results were obtained manually and were checked using the {\sl Maple} computer-algebra program.

\begin{eqnarray}
{\bf M^2}\,Y_\frac{1}{2}^\frac{1}{2} = {\bf M^2}\,e^{(\frac{1}{2}\,i\,\phi)}\,\sin^\frac{1}{2}\!\theta
&=& {\textstyle\frac{3}{4}}\,\hbar^2\,e^{(\frac{1}{2}\,i\,\phi)}\,(1\!-\!\cos^2\!\theta)/\sin^\frac{3}{2}\!\theta
\\\nonumber
&=& {\textstyle\frac{3}{4}}\,\hbar^2\,e^{(\frac{1}{2}\,i\,\phi)}\,\sin^\frac{1}{2}\!\theta =            {\textstyle\frac{1}{2}}\!\left({\textstyle\frac{1}{2}}\!+\!1\!\right)\,\hbar^2\,Y_\frac{1}{2}^\frac{1}{2}
\end{eqnarray}

\begin{eqnarray}
{\bf M^2}\,Y_\frac{3}{2}^\frac{3}{2} = {\bf M^2}\,e^{(\frac{3}{2}\,i\,\phi)}\,\sin^\frac{3}{2}\!\theta &=&
{\textstyle\frac{15}{4}}\,\hbar^2\,e^{(\frac{3}{2}\,i\,\phi)}\,(1\!-\!\cos^2\!\theta)/\sin^\frac{1}{2}\!\theta
\\\nonumber &=&
{\textstyle\frac{15}{4}}\,\hbar^2\,e^{(\frac{3}{2}\,i\,\phi)}\,\sin^\frac{3}{2}\!\theta =   {\textstyle\frac{3}{2}}\!\left({\textstyle\frac{3}{2}}\!+\!1\!\right)\,\hbar^2\,Y_\frac{3}{2}^\frac{3}{2}
\end{eqnarray}

\begin{eqnarray}
{\bf M^2}\,Y_\frac{3}{2}^\frac{1}{2} = {\bf M^2}\,\cos\theta\,\sin^\frac{1}{2}\!\theta\,e^{(\frac{1}{2}\,i\,\phi)}
 &=&
{\textstyle\frac{15}{4}}\,\hbar^2\,\cos\theta\,e^{(\frac{1}{2}\,i\,\phi)}\,(1\!-\!\cos^2\!\theta)/
\sin^\frac{3}{2}\!\theta
\\\nonumber &=&
{\textstyle\frac{15}{4}}\,\hbar^2\,\cos\theta\,\sin^\frac{1}{2}\!\theta\,e^{(\frac{1}{2}\,i\,\phi)} =
{\textstyle\frac{3}{2}}\!\left({\textstyle\frac{3}{2}}\!+\!1\!\right)\,\hbar^2\,Y_\frac{3}{2}^\frac{1}{2}
\end{eqnarray}

\begin{eqnarray}
{\bf M^2}Y_\frac{5}{2}^\frac{5}{2} = {\bf M^2}\,\sin^\frac{5}{2}\!\theta\,e^{(\frac{5}{2}\,i\,\phi)} &=&
{\textstyle\frac{35}{4}}\,\hbar^2\,\sin^\frac{1}{2}\!\theta\,e^{(\frac{5}{2}\,I\,\phi)}\,(1\!-\!\cos^2\!\theta)
\\\nonumber &=&
{\textstyle\frac{35}{4}}\,\hbar^2\,\sin^\frac{5}{2}\!\theta\,e^{(\frac{5}{2}\,i\,\phi)} =   {\textstyle\frac{5}{2}}\!\left({\textstyle\frac{5}{2}}\!+\!1\!\right)\,\hbar^2\,Y_\frac{5}{2}^\frac{5}{2}
\end{eqnarray}

\begin{eqnarray}
{\bf M^2}Y_\frac{5}{2}^\frac{3}{2} = {\bf M^2}\,\cos\theta\,\sin^\frac{3}{2}\!\theta\,e^{(\frac{3}{2}\,i\,\phi)}
 &=&
{\textstyle\frac{35}{4}}\,\hbar^2\,\cos\theta\,e^{(\frac{3}{2}\,I\,\phi)}\,(1\!-\!\cos^2\!\theta)/
\sin^\frac{1}{2}\!\theta \\\nonumber &=&
{\textstyle\frac{35}{4}}\,\hbar^2\,\cos\theta\,\sin^\frac{3}{2}\!\theta\,e^{(\frac{3}{2}\,i\,\phi)} =  {\textstyle\frac{5}{2}}\!\left({\textstyle\frac{5}{2}}\!+\!1\!\right)\,\hbar^2\,Y_\frac{5}{2}^\frac{3}{2}
\end{eqnarray}

\begin{eqnarray}\nonumber
{\bf M^2}Y_\frac{5}{2}^{\frac{1}{2}} =
{\bf M^2}\,e^{(\frac{1}{2}\,i\,\phi}\,\sin^\frac{1}{2}\!\theta\,(1\!-\!4\cos^2\!\theta) &=&
{\textstyle\frac{35}{4}}\,\hbar^2\,e^{(\frac{1}{2}\,i\,\phi)}\,(1\!-\!5\cos^2\!\theta+4\cos^4\!\theta)/
\sin^\frac{3}{2}\!\theta\\\nonumber &=&
{\textstyle\frac{35}{4}}\,\hbar^2\,e^{(\frac{1}{2}\,i\,\phi)}\,(1\!-\!4\cos^2\!\theta)(1\!-\!\cos^2\!\theta)/
\sin^\frac{3}{2}\!\theta\\\nonumber &=&
{\textstyle\frac{35}{4}}\,\hbar^2\,e^{(\frac{1}{2}\,i\,\phi)}\,\sin^\frac{1}{2}\!\theta\,(1\!-\!4\cos^2\!\theta) =   {\textstyle\frac{5}{2}}\!\left({\textstyle\frac{5}{2}}\!+\!1\!\right)\,\hbar^2\,Y_\frac{5}{2}^{\frac{1}{2}}\\
\end{eqnarray}

\filbreak
\subsubsection{Some General Cases: $\ell=|m|$, $\ell=|m|\!+\!1$, $\ell=|m|\!+\!2$ and $\ell=|m|\!+\!3$}
\paragraph{In the case $\ell=|m|$}$\!\!\!$Table 1 of \cite{Hunteretal}  (the column for $i\!=\!0$)
shows that the polynomial $P_\ell^{|m|}$
is simply a constant (equal to 1 in Table 1 of \cite{Hunteretal}) and
hence the spherical harmonic has the simple form:
\begin{equation}\label{LeqM}
Y_\ell^{m} = e^{i\,m\,\phi}(\sin\theta)^{|m|}
\end{equation}
From this simple form it follows that $Y_\ell^{m}$ is an
eigenfunction of ${\bf M^2}$ for all values of $\ell$:
\begin{eqnarray}
{\bf M^2}\,Y_\ell^{m} = {\bf M^2}\,e^{(i\,m\,\phi)}\,(\sin\theta)^{|m|}
= {|m|}(|m|\!+\!1\!)\,\hbar^2\,Y_\ell^{m}
= {\ell}(\ell\!+\!1\!)\,\hbar^2\,Y_\ell^{m}
\end{eqnarray}

\paragraph{In the case $\ell=|m|+1$}$\!\!\!$ Table 1 of \cite{Hunteretal}  (the column for $i\,$=$\,1$)
shows that the polynomial $P_\ell^{|m|}$
is simply $\cos\theta$ and hence the spherical harmonic has the form:
\begin{equation}\label{LeqMp1}
Y_\ell^{m} = e^{i\,m\,\phi}(\sin\theta)^{|m|}\,\cos\theta
\end{equation}
From this simple form Maple deduced that $Y_\ell^{m}$ is an
eigenfunction of ${\bf M^2}$ for all values of $m$:
\begin{eqnarray}
{\bf M^2}\,Y_\ell^{m} = {\bf M^2}\,e^{(i\,m\,\phi)}\,(\sin\theta)^{|m|}\cos\theta
= (|m|\!+\!1)(|m|\!+\!2\!)\,\hbar^2\,Y_\ell^{m}
= {\ell}(\ell\!+\!1\!)\,\hbar^2\,Y_\ell^{m}
\end{eqnarray}

\paragraph{In the case $\ell=|m|+2$}$\!\!\!$Table 1 of \cite{Hunteretal} (the column for $i\,$=$\,2$) shows that  the polynomial $P_\ell^{|m|}$
is  $\left[1\!-\!(2\,m+3)\cos^2\!\theta\right]$ and
hence the spherical harmonic has the form:
\begin{equation}\label{LeqMp2}
Y_\ell^{m} = e^{i\,m\,\phi}(\sin\theta)^{|m|}\,\left[1\!-\!(2\,m+3)\cos^2\!\theta\right]
\end{equation}
From this form Maple proved that $Y_\ell^{m}$ is an
eigenfunction of ${\bf M^2}$ for all values of $m$:
\begin{eqnarray}
{\bf M^2}\,Y_\ell^{m} &=&
{\bf M^2}\,e^{(i\,m\,\phi)}\,(\sin\theta)^{|m|}\left[1\!-\!(2\,m+3)\cos^2\!\theta\right]\\
&=& (|m|\!+\!2)(|m|\!+\!3\!)\,\hbar^2\,Y_\ell^{m}
= {\ell}(\ell\!+\!1\!)\,\hbar^2\,Y_\ell^{m}
\end{eqnarray}

\paragraph{In the case $\ell=|m|+3$}$\!\!\!$Table 1 of \cite{Hunteretal} (the column for $i\,$=$\,3$) shows that  the polynomial $P_\ell^{|m|}$
is  $\cos\theta\left[3\!-\!(2\,m\!+\!5)\cos^2\!\theta\right]$ and
hence the spherical harmonic has the form:
\begin{equation}\label{LeqMp3}
Y_\ell^{m} = e^{i\,m\,\phi}(\sin\theta)^{|m|}\,\cos\theta\left[3\!-\!(2\,m\!+\!5)\cos^2\!\theta\right]
\end{equation}
From this form Maple proved that $Y_\ell^{m}$ is an
eigenfunction of ${\bf M^2}$ for all values of $m$:
\begin{eqnarray}
{\bf M^2}\,Y_\ell^{m} &=&
{\bf M^2}\,e^{(i\,m\,\phi)}\,(\sin\theta)^{|m|}\cos\theta\left[3\!-\!(2\,m\!+\!5)\cos^2\theta\right]\\
&=& (|m|\!+\!3)(|m|\!+\!4\!)\,\hbar^2\,Y_\ell^{m}
= {\ell}(\ell\!+\!1\!)\,\hbar^2\,Y_\ell^{m}
\end{eqnarray}

\subsubsection{Summary for ${\bf M^2}$}
The above results prove that $Y_\ell^{m}$ is an eigenfunction of  ${\bf M^2}$ when
$\ell\!=\!|m|$, $\ell\!=\!|m|\!+\!1$, $\ell\!=\!|m|\!+\!2$ and $\ell\!=\!|m|\!+\!3$, for all values of $\ell$;
i.e. for all the Legendre Functions in Table (1 of \cite{Hunteretal}) except those in the last two columns,
{\em for all values} of $|m|$ from 0 to $\infty$ {\em including the half-odd-integer values};
i.e.~for all rows of Table 1 of \cite{Hunteretal} extended to $|m|\!\rightarrow\!\infty$.

We didn't prove any more results explicitly, but it is a plausible induction from the proven results
that $Y_\ell^{m}$ is always an eigenfunction of  ${\bf M^2}$  for all values of $\ell$ and $m$ including
the functions for which $\ell$ and $m$ have half-odd-integer values.

\section{Ladder Operations}

Since the ladder operators, ${\bf M_+}$ and ${\bf M_-}$, normally transform $Y_\ell^m$ into
 $Y_\ell^{m\!+\!1}$ and $Y_\ell^{m\!-\!1}$ respectively, it is appropriate to work with the operators
defined by (\ref{ladderopP}) \underline{divided b}y\underline{ $\hbar$} in order to avoid multiplying the
result of the transformation by a factor of $\hbar$ ; these renormalized ladder operators are
distinguished by primes, ${\bf {M'}_+}$ and ${\bf {M'}_-}$:
\begin{eqnarray}\label{renormMPM}
{\bf {M'}_+} = {\bf M_+}/\hbar =
e^{(i\,\phi)}\!\left(\frac{\partial}{\partial\theta} +i\,\cot\theta \,\frac{\partial}{\partial\phi}\right)\quad\quad
{\bf {M'}_-}  = {\bf M_-}/\hbar =
e^{(-i\,\phi)}\!\left(\frac{\partial}{\partial\theta} - i\,\cot\theta \,\frac{\partial}{\partial\phi}\right)
\end{eqnarray}

\subsection{$\phi$ Dependence}
A general conclusion about operating on a spherical harmonic, $Y_\ell^m$, with the ladder operators,
${\bf {M'}_+}$ and ${\bf {M'}_-}$, is that the $\phi$ dependence of the result is always the normal, expected
result for both integer and half-odd integer harmonics.  This conclusion is inferred by observing
that the differentiation w.r.t.~$\phi$ in ${\bf {M'}_+}$ and ${\bf {M'}_-}$ leaves the $\phi$ dependence of
$Y_\ell^m$ unchanged because the $\phi$ dependence of $Y_\ell^m$ is the exponential function
$\exp\{i\,m\,\phi\}$.
In addition, ${\bf {M'}_+}$ and ${\bf {M'}_-}$ multiply $Y_\ell^m$ by
$\exp(i\,\phi)$ and $\exp(\!-\!i\,\phi)$ respectively; this multiplication has the effect of
increasing and decreasing the exponent by 1 respectively; i.e.:
\begin{equation}
{\bf {M'}_+}\,e^{\{i\,m\,\phi\}} = -m\,\cot\theta\,e^{\{i(m\!+\!1)\phi\}}\quad\quad
{\bf {M'}_-}\,e^{\{i\,m\,\phi\}} =  +m\,\cot\theta\,e^{\{i(m\!-\!1)\phi\}}
\end{equation}

\subsection{$\theta$ Dependence}
The differentiation w.r.t.~$\theta$ in ${\bf {M'}_+}$ and ${\bf {M'}_-}$ of
$\sin^{|m|}\!\theta\,P_\ell^{|m|}$ will produce:
\begin{eqnarray}
\frac{d }{d\,\theta}\sin^{|m|}\!\theta\,P_\ell^{|m|} =
\sin^{|m|}\!\theta\left\{|m|\,\frac{\cos\theta}{\sin\theta}\,P_\ell^{|m|} +
\frac{d P_\ell^{|m|}}{d \theta}\right\}
\end{eqnarray}
Recognition of $P_\ell^{|m|}$ as a polynomial in $x\!=\!\cos\theta$ (as shown in Table 1 of \cite{Hunteretal})
suggests that some simplification will result from expressing the differentiation of  $P_\ell^{|m|}$
w.r.t. $x$ rather than $\theta$, since:
\vspace{-\baselineskip}
\begin{eqnarray}
\frac{d }{d \theta}\,P_\ell^{|m|} = {-\sin\theta}\,\frac{d P_\ell^{|m|}}{d x}
\end{eqnarray}
Thus the general results are:
\begin{eqnarray}\nonumber
{\bf {M'}_+}\,Y_\ell^m &=&
{\bf {M'}_+}\,\left\{e^{(i\,m\,\phi)}\,sin^{|m|}\theta\,P_\ell^{|m|}\right\}
\\\label{MPoui}
&=& e^{(i\,[m\!+\!1]\phi)}
\sin^{(|m|+1)}\!\theta\left\{(|m|\!-\!m)\,\frac{\cos\theta}{\sin^2\!\theta}\,P_\ell^{|m|} -
\frac{d P_\ell^{|m|}}{d x}\right\}\\\label{MPnon}
&=& e^{(i\,[m\!+\!1]\phi)}
\sin^{(|m|-1)}\!\theta\left\{(|m|\!-\!m)\,\cos\theta\,P_\ell^{|m|} -
{\sin^2\!\theta}\,\frac{d  P_\ell^{|m|}}{d  x}\right\}
\end{eqnarray}
\begin{eqnarray}\nonumber
{\bf {M'}_-}\,Y_\ell^m &=&
{\bf {M'}_-}\,\left\{e^{(i\,m\,\phi)}\,sin^{|m|}\theta\,P_\ell^{|m|}\right\}
\\\label{MMnon}
&=& e^{(i\,[m\!-\!1]\phi)}
\sin^{(|m|+1)}\!\theta\left\{(|m|\!+\!m)\,\frac{\cos\theta}{\sin^2\!\theta}\,P_\ell^{|m|} -
\frac{d P_\ell^{|m|}}{d x}\right\}\\\label{MMoui}
&=& e^{(i\,[m\!-\!1]\phi)}
\sin^{(|m|-1)}\!\theta\left\{(|m|\!+\!m)\,\cos\theta\,P_\ell^{|m|} -
{\sin^2\!\theta}\,\frac{d  P_\ell^{|m|}}{d  x}\right\}
\end{eqnarray}
Which of the two alternative, equivalent expressions for ${\bf {M'}_+}\,Y_\ell^m$ (\ref{MPoui},\ref{MPnon}) and the
two for ${\bf {M'}_-}\,Y_\ell^m$ (\ref{MMnon},\ref{MMoui})
produces the simpler result depends upon whether $m$ is positive or negative;
the term in (\ref{MPoui}) and in (\ref{MPnon}) having a factor of $(|m|\!-\!m)$ will be zero when $m$ is positive;
likewise the term in (\ref{MMnon}) and in (\ref{MMoui}) having a factor of $(|m|\!+\!m)$ will be zero when $m$ is negative.

\subsubsection{The Case of Positive $m$}
In the cases where ${\bf {M'}_+}$ and ${\bf {M'}_-}$ operate on a $Y_\ell^m$ with $m$ positive,
$m\!=\!|m|$,
and hence the expressions (\ref{MPoui}) and (\ref{MMoui}) produce simpler results than (\ref{MPnon}) and (\ref{MMnon}) respectively:
\begin{eqnarray}\label{MPmP}
{\bf {M'}_+}\,Y_\ell^m
&=& - e^{(i\,[m\!+\!1]\phi)}
\sin^{(|m|+1)}\!\theta\left\{\frac{d P_\ell^{|m|}}{d x}\right\}
\end{eqnarray}
\begin{eqnarray}\label{MMmP}
{\bf {M'}_-}\,Y_\ell^m
&=& e^{(i\,[m\!-\!1]\phi)}
\sin^{(|m|-1)}\!\theta\left\{2\,m\,\cos\theta\,P_\ell^{|m|} -
{\sin^2\!\theta}\frac{d  P_\ell^{|m|}}{d  x}\right\}
\end{eqnarray}

\subsubsection{The Case of Negative $m$}
In the cases where ${\bf {M'}_+}$ and ${\bf {M'}_-}$ operate on a $Y_\ell^m$ with $m$ negative,
$m\!=\!-|m|$,
and hence (\ref{MPnon}) and (\ref{MMnon}) produce simpler results than (\ref{MPoui}) and (\ref{MMoui}) respectively:
\begin{eqnarray}\label{MPmN}
{\bf {M'}_+}\,Y_\ell^m
&=& e^{(i\,[m\!+\!1]\phi)}
\sin^{(|m|-1)}\!\theta\left\{2|m|\,{\cos\theta}\,P_\ell^{|m|} - {{\sin^2\!\theta}\,\frac{d P_\ell^{|m|}}{d x}}\right\}
\end{eqnarray}
\begin{eqnarray}\label{MMmN}
{\bf {M'}_-}\,Y_\ell^m
&=& - e^{(i\,[m\!-\!1]\phi)}\,\sin^{(|m|\!+\!1)}\theta\left\{\frac{d  P_\ell^{|m|}}{d  x}\right\}
\end{eqnarray}

\subsubsection{The Case $\ell=|m|$}
In this case the polynomial $P_\ell^{|m|}$ is simply a
constant (the column for $i\!=\!0$ in Table 1 of \cite{Hunteretal}), and since
the spherical harmonic has the simple form of (\ref{LeqM}) we can derive the effect of operating
with ${\bf {M'}_+}$ and ${\bf {M'}_-}$ for all values of $m$; the derivatives
${d P_\ell^{|m|}}\!/{d x}$ in (\ref{MPmP},\ref{MMmP},\ref{MPmN},\ref{MMmN}) are zero, and hence
from the fact that (\ref{MPmP}) and (\ref{MMmN}) both have this zero derivative as a factor  it follows that:
\vspace{-\baselineskip}
\begin{eqnarray}\label{MPmPmeqL}
{\bf {M'}_+}\,Y_\ell^\ell = 0\quad\quad{\rm and}\quad\quad
\label{MMmNmeqL}
{\bf {M'}_-}\,Y_\ell^{-\ell} = 0
\end{eqnarray}
which are the usual results that the ladder operators produce a zero result (anihilation) when
${\bf {M'}_+}$ operates on $\,Y_\ell^\ell$, and when ${\bf {M'}_-}$ operates on $\,Y_\ell^{-\ell}$.

The other two results for  $\ell=|m|$ are obtained from (\ref{MMmP}) and (\ref{MPmN}) respectively:
\begin{eqnarray}\label{MMmPmeqL}
{\bf {M'}_-}\,Y_\ell^\ell
&=& e^{[i\,(\ell\!-\!1)\phi]}\,
\sin^{(\ell\!-\!1)}\!\theta\,\left\{2\,\ell\,\cos\theta\right\} \\\nonumber
&=& 2\,\ell\,Y_\ell^{(\ell\!-\!1)}\quad{\rm if}\; \ell\!\ge\!1
\end{eqnarray}
\begin{eqnarray}\label{MPmNmeqL}
{\bf {M'}_+}\,Y_\ell^{-\ell}
&=& e^{[-i\,(\ell\!-\!1)\phi]}\,
\sin^{(\ell\!-\!1)}\!\theta\,\left\{2\ell\,{\cos\theta}\right\} \\\nonumber
&=& 2\,\ell\,Y_\ell^{(-\ell\!+\!1)}\quad{\rm if}\; \ell\!\ge\!1
\end{eqnarray}
These are the expected results as long as  $\ell\!\ge\!1$;  i.e.~the functions in the column for $i\!\!=\!\!1$ in Table 1 of \cite{Hunteretal}
excluding the row for $|m|\!=\!\frac{1}{2}$.\footnote{In the case of the first row of Table 1 of \cite{Hunteretal} (the case $|m|=0$)
equations (\ref{MMmPmeqL}) and (\ref{MPmNmeqL}) produce the expected anihilation results
because of the factor of $\ell=0$ on their right hand sides.}
The exceptional cases are:
\begin{eqnarray}\label{MMmPmeqL12}
{\bf {M'}_-}\,Y_\frac{1}{2}^\frac{1}{2}
= e^{(-i\frac{\phi}{2})}\,
\sin^{-\frac{1}{2}}\!\theta\,\left\{2\,\ell\,\cos\theta\right\} &=& Y_\frac{1}{2}^{-\frac{1}{2}}\!\times\!\cot\theta \neq {\rm const}\!\times\! Y_\frac{1}{2}^{-\frac{1}{2}}
\\\label{MPmNmeqL12}
{\bf {M'}_+}\,Y_\frac{1}{2}^{-\frac{1}{2}}
= e^{(i\frac{\phi}{2})}
\sin^{-\frac{1}{2}}\!\theta\,\left\{2\,\ell\,\cos\theta\right\} &=& Y_\frac{1}{2}^{\frac{1}{2}}\!\times\!\cot\theta \neq {\rm const}\!\times\! Y_\frac{1}{2}^{\frac{1}{2}}
\end{eqnarray}
These exceptional results for $\ell=|m|=\frac{1}{2}$ have a $\theta$ factor of $\cos\theta/\sqrt{\sin\theta}$,
whereas in the expected result this factor would be
$\sqrt{\sin\theta}$.

\subsection{Cases: $\ell\!=\!\frac{1}{2}$, $\ell\!=\!\frac{3}{2}$ and $\ell\!=\!\frac{5}{2}$}

Operation on the 12 half-odd integer functions defined in (\ref{Y1212})-(\ref{Y1252})
with each of ${\bf {M'}_+}$ and ${\bf {M'}_-}$ defined in (\ref{renormMPM}) produced the following results;
these results were (like the eigenvalue results) obtained manually and
then checked using Maple computer algebra.\footnote{some of the Maple results were simplified manually.}

\subsubsection{Cases for $\ell\!=\!\frac{1}{2}$}
The results of applying the ladder operators  ${\bf {M'}_+}$ and ${\bf {M'}_-}$ [i.e.~(\ref{renormMPM})] to the two spherical harmonics
for $\ell\!=\!\frac{1}{2}$ [i.e.~(\ref{Y1212})] are explicated by
equations (\ref{MPmPmeqL}), (\ref{MMmPmeqL12}) and (\ref{MPmNmeqL12}) above.\footnote{since in all four of these cases
$\ell\!=\!|m|$.}
\paragraph{The first two results (\ref{MPmPmeqL}) for ${\bf \ell=\frac{1}{2}}$ show:}that Merzbacher's
inference \cite[p.241,col.2]{MerzbacherAJP}
 that ``the ladder does not terminate''  is incorrect; he made this inference
by operating with {\em the square} of ${\bf {M'}_-}$ on $Y_\frac{1}{2}^{\frac{1}{2}}$:
\vspace{-\baselineskip}
\begin{eqnarray}
\left({\bf {M'}_-}\right)^2Y_\frac{1}{2}^{\frac{1}{2}} \ne  0
\end{eqnarray}
which is true because in the {\em first} application of ${\bf {M'}_-}$ on  $Y_\frac{1}{2}^{\frac{1}{2}}$ the result is the abnormal (\ref{MMmPmeqL12}),
and the {\em second} application of ${\bf {M'}_-}$ on this abnormal result is indeed not zero.
However, Merzbacher is incorrect in inferring from this result (via the abnormal intermediate result (\ref{MMmPmeqL12}))
 that ``the ladder does not terminate'' \cite[p.241,col.2]{MerzbacherAJP};
equations (\ref{MPmPmeqL}) demonstrate that the ladder does indeed terminate
as it is expected to do.  Fortissimo, this general result demonstrates that the ladder terminates {\em at both ends}
for all values of $\ell$, whereas Merzbacher was only concerned with the case $\ell=\frac{1}{2}$.

\subsubsection{Cases for $\ell\!=\!\frac{3}{2}$}
Application of ${\bf {M'}_+}$ and ${\bf {M'}_-}$ [defined by (\ref{renormMPM})] to the two spherical harmonics
defined by (\ref{Y3232}) produces four results:
\vspace{-\baselineskip}
\begin{eqnarray}
{\bf {M'}_+}Y_\frac{3}{2}^{\frac{3}{2}} = 0\quad\quad\quad
{\bf {M'}_-} Y_\frac{3}{2}^{-\frac{3}{2}} = 0
\end{eqnarray}
which are instances of the general anihilation results (\ref{MPmPmeqL}), and
\begin{eqnarray}
{\bf {M'}_-} Y_\frac{3}{2}^{\frac{3}{2}} &=& 3\,e^{(\frac{1}{2}\,i\,\phi)}\,\sqrt{\sin\theta}\,\cos\theta
 = 3\,Y_\frac{3}{2}^{\frac{1}{2}}
\end{eqnarray}
\begin{eqnarray}
{\bf {M'}_+}Y_\frac{3}{2}^{-\frac{3}{2}} &=&
3\,e^{(-\frac{1}{2}\,i\,\phi)}\,\sqrt{\sin\theta}\,\cos\theta = 3\,Y_\frac{3}{2}^{-\frac{1}{2}}
\end{eqnarray}
which are instances of the generally expected results (\ref{MMmPmeqL}) and (\ref{MPmNmeqL}) for $\ell\!\ge\!1$.

Application of ${\bf {M'}_+}$ and ${\bf {M'}_-}$  to the two spherical harmonics
defined by (\ref{Y1232}) produces four results:
\vspace{-\baselineskip}
\begin{eqnarray}
{\bf {M'}_+}Y_\frac{3}{2}^{\frac{1}{2}} &=&
- e^{(3/2\,i\,\phi)}\,\sin\theta^{(3/2)}
 = -Y_\frac{3}{2}^{\frac{3}{2}}
\end{eqnarray}
\begin{eqnarray}
{\bf {M'}_-} Y_\frac{3}{2}^{-\frac{1}{2}} &=&  - e^{(-3/2\,i\,\phi)}\,\sin\theta^{(3/2)}
 = -Y_\frac{3}{2}^{-\frac{3}{2}}
\end{eqnarray}
which are instances of the usually expected results, and
\begin{eqnarray}\label{UNexpectP3}
{\bf {M'}_-} Y_\frac{3}{2}^{\frac{1}{2}} &=& e^{(-\frac{1}{2}\,i\,\phi)}\,
\frac{(2\,\cos^2\!\theta\!-\!1)}{\sqrt{\sin\theta}}
 = 2\cot(2\theta)\, Y_\frac{3}{2}^{-\frac{1}{2}}
\neq{\rm const}\!\times\! Y_\frac{3}{2}^{-\frac{1}{2}}
\end{eqnarray}
\begin{eqnarray}\label{UNexpectM3}
{\bf {M'}_+} Y_\frac{3}{2}^{-\frac{1}{2}} &=& e^{(\frac{1}{2}\,i\,\phi)}\,
\frac{(2\,\cos^2\!\theta\!-\!1)}{\sqrt{\sin\theta}}
 = 2\cot(2\theta)\, Y_\frac{3}{2}^{\frac{1}{2}}
\neq{\rm const}\!\times\! Y_\frac{3}{2}^{\frac{1}{2}}
\end{eqnarray}
which are {\em not} the usually
expected results because the multiplier of the expected function is not a constant; i.e.
\vspace{-\baselineskip}
\begin{eqnarray}
Y_\frac{3}{2}^{\pm\frac{1}{2}} = e^{(\pm\frac{1}{2}\,i\,\phi)}\,{\sqrt{\sin\!\theta}}\,\cos\!\theta
\end{eqnarray}
from Table 1 of \cite{Hunteretal} (the entry for $|m|\!=\!\frac{1}{2}$, $i\!=\!1$).  Results  (\ref{UNexpectP3}) and (\ref{UNexpectM3})
are distinct from the $\ell\!=\!|m|$ results of (\ref{MMmPmeqL}) and (\ref{MPmNmeqL}).

\subsubsection{Cases for $\ell\!=\!\frac{5}{2}$}
Application of ${\bf {M'}_+}$ and ${\bf {M'}_-}$ to the two spherical harmonics
defined by (\ref{Y5252}) produces four results:
\vspace{-\baselineskip}
\begin{eqnarray}
{\bf {M'}_+}Y_\frac{5}{2}^{\frac{5}{2}} = 0\quad\quad\quad
{\bf {M'}_-}Y_\frac{5}{2}^{-\frac{5}{2}} = 0
\end{eqnarray}
which are instances of the general anihilation results (\ref{MPmPmeqL}), and
\begin{eqnarray}
{\bf {M'}_-} Y_\frac{5}{2}^{\frac{5}{2}}
= 5\,e^{(3/2\,i\,\phi)}\,\sin^{(3/2)}\!\theta\,\cos\theta = 5 Y_\frac{5}{2}^{\frac{3}{2}}
\end{eqnarray}
\begin{eqnarray}
{\bf {M'}_+}Y_\frac{5}{2}^{-\frac{5}{2}}
= 5\,e^{-(3/2\,i\,\phi)}\,\sin^{(3/2)}\!\theta\,\cos\theta = 5 Y_\frac{5}{2}^{-\frac{3}{2}}
\end{eqnarray}
which are instances of the general results (\ref{MMmPmeqL}) and (\ref{MPmNmeqL}) for $\ell\!\ge\!1$.

Application of ${\bf {M'}_+}$ and ${\bf {M'}_-}$ to the two spherical harmonics
defined by (\ref{Y3252}) produces four results:
\vspace{-\baselineskip}
\begin{eqnarray}
{\bf {M'}_+}Y_\frac{5}{2}^{\frac{3}{2}} =   - e^{(5/2\,i\,\phi)}\,\sin\theta^{(5/2)}
= - Y_\frac{5}{2}^{\frac{5}{2}}
\end{eqnarray}
\begin{eqnarray}
{\bf {M'}_-} Y_\frac{5}{2}^{-\frac{3}{2}} =  - e^{(-5/2\,i\,\phi)}\,\sin\theta^{(5/2)}
= - Y_\frac{5}{2}^{-\frac{5}{2}}
\end{eqnarray}
which are instances of the usually expected results, and
\begin{eqnarray}
{\bf {M'}_-} Y_\frac{5}{2}^{\frac{3}{2}} =
e^{(\frac{1}{2}\,i\,\phi)}\,\sqrt{\sin\theta}\,(4\,\cos\theta^{2} - 1)
= - Y_\frac{5}{2}^{\frac{1}{2}}
\end{eqnarray}
\begin{eqnarray}
{\bf {M'}_+} Y_\frac{5}{2}^{-\frac{3}{2}} =
e^{(-\frac{1}{2}\,i\,\phi)}\,\sqrt{\sin\theta}\,(4\,\cos\theta^{2} - 1)
= - Y_\frac{5}{2}^{-\frac{1}{2}}
\end{eqnarray}
which are also instances of the usually expected results.

Application of ${\bf {M'}_+}$ and ${\bf {M'}_-}$ to the two spherical harmonics
defined by (\ref{Y1252}) produces four results:
\vspace{-\baselineskip}
\begin{eqnarray}
{\bf {M'}_+}Y_\frac{5}{2}^{\frac{1}{2}} = 8\,\sin^{(3/2)}\!\theta\,\cos\theta\,e^{(3/2\,i\,\phi)}
= 8 Y_\frac{5}{2}^{\frac{3}{2}}
\end{eqnarray}
\begin{eqnarray}
{\bf {M'}_-} Y_\frac{5}{2}^{-\frac{1}{2}} = 8\,\sin\theta^{(3/2)}\,\cos\theta\,e^{(-3/2\,i\,\phi)} =
8 Y_\frac{5}{2}^{-\frac{3}{2}}
\end{eqnarray}
which are also instances of the usually expected results, and
\begin{eqnarray}\nonumber
{\bf {M'}_-} Y_\frac{5}{2}^{\frac{1}{2}} &=&
- 3\,e^{(-\frac{1}{2}\,i\,\phi)}\,
\frac{\cos\!\theta(4\,\cos^2\!\theta-3)}{\sqrt{\sin\!\theta}} \\\nonumber\\\label{UNexpectP5}
 &=&
- 3\,e^{(-\frac{1}{2}\,i\,\phi)}\,{\sqrt{\sin\!\theta}}\,
\frac{\cos(3\theta)}{\sin\!\theta} =
3 \cot(3\theta)\, Y_\frac{5}{2}^{-\frac{1}{2}}
\neq{\rm const}\!\times\! Y_\frac{5}{2}^{-\frac{1}{2}}
\end{eqnarray}
\begin{eqnarray}\nonumber
{\bf {M'}_+} Y_\frac{5}{2}^{-\frac{1}{2}} &=&
- 3\,e^{(\frac{1}{2}\,i\,\phi)}\,
\frac{\cos\!\theta(4\,\cos^2\!\theta-3)}{\sqrt{\sin\!\theta}}
\\\nonumber\\\label{UNexpectM5}
&=&
- 3\,e^{(\frac{1}{2}\,i\,\phi)}\,{\sqrt{\sin\!\theta}}\,
\frac{\cos(3\theta)}{\sin\!\theta}
= 3\cot(3\theta)\, Y_\frac{5}{2}^{\frac{1}{2}}
\neq{\rm const}\!\times\! Y_\frac{5}{2}^{\frac{1}{2}}
\end{eqnarray}
since:\footnote{from Table 1 of \cite{Hunteretal} (the entry for $|m|\!=\!\frac{1}{2}$, $i\!=\!2$)}
\vspace{-\baselineskip}
\begin{eqnarray}
Y_\frac{5}{2}^{\pm\frac{1}{2}} = e^{(\pm\frac{1}{2}\,i\,\phi)}\,{\sqrt{\sin\!\theta}}\,(1-4\cos^2\!\theta)
\end{eqnarray}
These results, (\ref{UNexpectP5}) and (\ref{UNexpectM5}), are {\em not} the usually
expected results because the multiplier of the expected function is not a constant.

\subsection{Overall Conclusion}

\paragraph{The ladder operators produce the usually expected result for all values of $\ell$} except in two cases:
\vspace{-\baselineskip}
\begin{eqnarray}
{\bf {M'}_+} Y_\ell^{-\frac{1}{2}} \neq{\rm const}\!\times\! Y_\ell^{\frac{1}{2}}
\\\nonumber\\
{\bf {M'}_-} Y_\ell^{\frac{1}{2}} \neq{\rm const}\!\times\! Y_\ell^{-\frac{1}{2}}
\end{eqnarray}
From the computed instances of these unusual results
(\ref{MMmPmeqL12},\ref{MPmNmeqL12},\ref{UNexpectP3},\ref{UNexpectM3},\ref{UNexpectP5},\ref{UNexpectM5}) the
general non-constant multiplier appears to be:
\begin{eqnarray}\label{genMPhalf}
{\bf {M'}_+} Y_\ell^{-\frac{1}{2}} &=&
(\ell+\!{\textstyle\frac{1}{2}})\,{\cot[(\ell\!+\!{\textstyle\frac{1}{2}})\theta]}\,Y_\ell^{\frac{1}{2}}
\\\nonumber\\\label{genMMhalf}
{\bf {M'}_-} Y_\ell^{\frac{1}{2}} &=&
(\ell+\!{\textstyle\frac{1}{2}})\,{\cot[(\ell\!+\!{\textstyle\frac{1}{2}})\theta]}\,Y_\ell^{-\frac{1}{2}}
\end{eqnarray}
for all values of $\ell$, which are, of course, half-odd-integer values.

These failures in the Schr\"odinger represention
are not present in the abstract theory of angular momentum
based upon the fundamental commutation relations \cite[\S 5.4,pp.119-120]{Levine}.

The algebraic origin of the failures can be understood by reference to the general results
(\ref{MMmP}) and (\ref{MPmN}).   For the simplest case of $\ell\!=\!|m|$;
 $P_\ell^{|m|}$ is a constant, which makes ${d P_\ell^{|m|\!}}/{d \theta}\!=\!0$
in (\ref{MMmP}) and (\ref{MPmN}), but
$|m|$ multiplies the $\cos\theta$ in these formulae:
\begin{itemize}
\item For integer $\ell=m=1$ the $\cos\theta$ is exactly the expected, usual result,  $Y_1^{0}$, but
\item For $\ell=|m|=\frac{1}{2}$ the $\cos\theta$ multiplies the expected result,
$Y_\frac{1}{2}^{\pm\frac{1}{2}}$ divided by $\sin\theta$
\end{itemize}
which is the algebraic reason for the failure.


\section{Double-Valuedness}\label{DV}

A distinctive feature of the Fermion Spherical Harmonics is that they are double-valued functions of
$\phi$; this arises because:
\begin{equation}\label{phifact1}
\exp[{i\,{\textstyle\frac{n}{2}}\,(\phi+2\pi)}] = {\bf -} \exp[{i\,{\textstyle\frac{n}{2}}\,\phi}]
\end{equation}
with $n$ an odd integer.  The angle $\phi$ must transit two complete circles for the
wavefunction to return to its original value:
\begin{equation}\label{phifact2}
\exp[{i\,{\textstyle\frac{n}{2}}\,(\phi+4\pi)}] = {\bf +} \exp[{i\,{\textstyle\frac{n}{2}}\,\phi}]
\end{equation}
This property of Fermion wavefunctions is well known \cite{Icke}.   The double-valuedness of the
wavefunction nevertheless leaves the probability single-valued, because the probability is computed
as the product of the $\phi$ factor on the R.H.S. of (\ref{phifact2}) with its
own complex conjugate, and thus the imaginery exponent produces a probability that
is independent of the angle $\phi$ for all values of the exponent; i.e.~of $n$ in (\ref{phifact1}) and (\ref{phifact2}).

\section{The Validity of the Fermion Harmonics}

The existence of Spherical Harmonics having half-odd integer quantum numbers, $\ell$ and $m$,
has been known to some scientists for many years \cite{Pauli,MerzbacherBook}; these and other authors have been concerned with
finding theoretical reasons for only using the spherical harmonics having integer values of $\ell$ and $m$.
A related question is why a given physical system has angular momentum states that {\em all} have integer values of $\ell$ and $m$, while
other systems have states that {\em all} have half-odd-integer values of $\ell$ and $m$; a simple proof that a particle
cannot have {\em both} integer and half-odd-integer values of $\ell$ and $m$, has been given by Bohm \cite[p.389]{Bohm}. However,
Merzbacher \cite{MerzbacherAJP} has shown that the orbital angular momentum  of a rigid body may be quantized in half-odd-integer values of
$\ell$ and $m$.

Merzbacher \cite[p.174]{MerzbacherBook} taking the single-valuedness
of the wavefunction as axiomatic advanced the argument  that the eigenfunction\footnote{of ${\bf M^2}$ and ${\bf M_z}$}
$\sqrt{sin\theta}\exp({i\,\phi/2})$,
can be made single-valued by limiting
the range of $\phi$ to 0 to $2\pi$; he then notes [using (\ref{phifact1})] that with
this restriction the $\phi$ factor is discontinuous at $\phi\!=\!2\pi$, and is therefore not differentiable
at $\phi\!=\!2\pi$, which makes it invalid because valid wavefunctions must be differentiable everywhere.

Buchdal \cite{Buchdal} questions the validity of Merzbacher's argument as being
too restrictive, and he quotes Bohm's argument \cite[p.389-390]{Bohm}
that it is only physically observable quantities that must be single valued.
This accords with our discussion above [after
(\ref{phifact1}) and (\ref{phifact2})], that the double valuedness  nevertheless leaves the
probability singled-valued.

Buchdal quotes Pauli \cite{Pauli} and notes that Pauli's ``not entirely simple'' argument
 has been misrepresented.  However, it should be noted that dismissing
 Merzbacher's argument and being sceptical about Pauli's
``not entirely simple'' argument, are peripheral to Buchdal's main purpose,
which is to advance an alternative argument (to the single-valuedness of the wavefunction)
to infer that the {\em orbital angular momentum} quantum numbers must be integers.
   Thus Buchdal only
infers that the half-integral spherical harmonics are invalid representations of {\em orbital}
angular momentum.

Likewise, the note published by Gray \cite{Gray} is only concerned with why half-integral angular
momenta are invalid representations of the {\em orbital} angular momentum of a particle.
His argument is based upon that of Bohm \cite{Bohm}; that a system
cannot have {\em both} integer and half-integer quantum numbers,
and since $\ell\!\!=\!\!m\!\!=\!\!0$ is known to be physically valid for orbital angular momentum
(the non-rotating state), and since
the ladder operations (relating different eigenfunctions having the same $\ell$ but
different $m$) step through the eigenfunctions in integer steps of $m$, he concludes that
all the values (including zero) of $m$ (and therefore of $\ell$) must be integers; it is important
to realize that Gray proposes
this argument as an alternative to Merzbacher's (single-valuedness of the wavefunction) argument
for the {\em orbital} angular momentum of a particle; Gray is not concerned with spin angular
momentum.

Blatt and Weisskopf \cite[p.783]{BlattBook} quote the unpublished argument of Nordsieck that dismisses single-valuedness
as unnecessary because only probability densities and expectation values must be singled-valued; his argument is reinforced
by noting that double-valued wave functions are used in the theory of particles with spin.

\section{Summary and Conclusions}

Spectroscopic results indicate that orbital angular momentum eigenstates
always correspond to integer values of $\ell$, whereas the intrinsic
angular momentum (spin) eigenstates of elementary particles (notably the electron),
 correspond to half-integer values of $\ell$ and $m$.

This difference is rooted in the diference between the orbital rotation of a body (which is classical apart from quantization of the angular momentum
in integer multiples of $\hbar$), and the intrinsic rotation of an elementary particle whose nature is not really understood.
Attempts to construct coherent models of the electron have not yet
yielded a physical understanding of the nature of spin \cite{Tomonaga,MacGregor}.
Dahl's paper ``The Spinning Electron'' \cite{Dahl} promises (in its Abstract and Introduction)
to present a physical model as a ``3-dimensional rotor governed by relativistic quantum mechanics'',
but the body of the paper leaves the reader still groping for a tangible physical model
of what an electron is, and in particular what gives rise to its spin and associated magnetic moment.

Including the Fermion Spherical Harmonics (for $\ell\!=\!\frac{1}{2}$) as part of a coherent
model of the electron implies adopting an interpretation of the angles $\theta$ and $\phi$;
the obvious interpretation is that they represent the orientation in space of the
particle's intrinsic angular momentum and its magnetic moment vector.  This explicit description of the orientation is, of course,
limited by the uncertainty principle; i.e.~while the $z$-component is defined by the value of $m$ (to be $m\hbar$), the total
angular momentum is always larger ($\hbar\sqrt{\ell(\ell+1)}$), and hence the angular momentum has a component in the $x$-$y$ plane
whose direction in space remains undetermined.
The explicit angular description may allow for a more precise theoretical description of physical systems such as the Stern-Gerlach experiment
\cite[p.141-148]{Fano}.

The failure of the ladder operations in the cases where $m$ changes sign, is not, in itself,
a convincing argument that the Fermion Spherical Harmonics are not valid as eigenfunctions of spin (and total)
angular momentum.  They are (as we have explicitly shown here) eigenfunctions of $M_z$ and $M^2$ with the expected
eigenvalues in all cases, and they are normalizable as shown in our previous paper \cite[Table 2]{Hunteretal}.
The eigenfunction property is essential for a valid quantum mechanical state, whereas ladder operators relating states with different eigenvalues are
only known for a few physical systems;\footnote{well-known for angular momentum and the one-dimensional harmonic oscillator; also known for the
hydrogen atom and for the electronic motion in the $H_2^+$ ion.}
 that this coordinate representation of spin angular momentum differs from the
abstract theory of angular momentum in this respect, is an interesting curiosity worthy of further investigation.

That the double-valued factors of the Fermion Spherical Harmonics ($\exp\{i\,n\,\phi/2\}$) occur
in the accepted Dirac and Pauli theories of spin supports their validity; however, in these established theories the wavefunction has 4 and 2 components
respectively, which seems to reduce the validity question to whether a
{\em scalar} wavefunction (with a Fermion Spherical Harmonic factor) can be an acceptable representation of spin
angular momentum.

\filbreak

\end{document}